          [1999/09/30 1.0 (PWD)]
\documentclass[a4paper,twocolumn]{esapub}

\usepackage{natbib,graphicx}

\title{The INTEGRAL Core Observing Programme}
\author{Christoph Winkler\footnote{On behalf of the INTEGRAL Science 
Working Team: G. Vedrenne, V. Sch\"onfelder, P. Ubertini, F. Lebrun, N. Lund,
A. Gimenez, T. Courvoisier, N. Gehrels, S. Grebenev, W. Hermsen, G. Palumbo,
J. Paul, R. Sunyaev, B. Teegarden, C. Winkler.}}
\affil{Astrophysics Division, ESA-ESTEC, 2200 AG Noordwijk, The Netherlands}

\begin{document}


\maketitle

\begin{abstract}
The Core Programme (CP) of the INTEGRAL mission is defined as the portion
of the scientific observing programme covering the guaranteed time observations
for the PI collaborations and other members of the INTEGRAL Science Working
Team. During the first two years of the mission, the guaranteed observing time
uses 35\% (year 1) and 30\% (year 2) of the total annual observing 
time ($\sim$26.6$\times$10$^{6}$s per year).
The Core Programme consists of three elements: a deep exposure of the Galactic
central radian, regular (weekly) 
scans of the Galactic plane and pointed observations.
This paper describes the current status of the INTEGRAL Core Programme and
summarizes the key elements.
\end{abstract}
                                                    
\section{Introduction}

\begin{figure}[htp]
\centering
\includegraphics[width=1.0\linewidth]{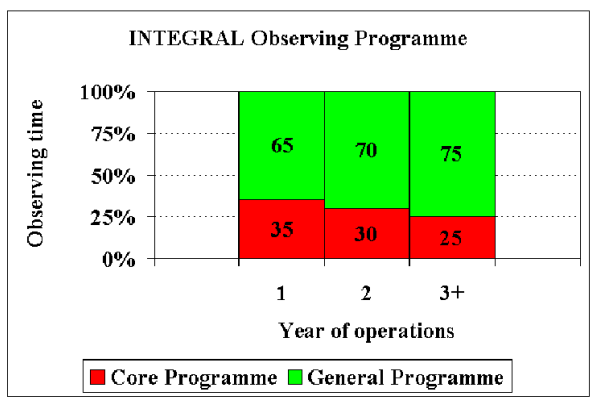}
\caption{Breakdown of INTEGRAL observing time.}
\label{cw:programme}
\end{figure}

The gamma-ray satellite INTEGRAL will be operated as an observatory
whith the majority of scientific observing time being awarded to
the scientific community at large, following standard AO and 
peer-review procedures:
this constitutes the General Programme covering open time observations.
As a return to those scientific collaborations and individual scientists
who contributed to the development, design and procurement of INTEGRAL
and who are represented in the INTEGRAL Science Working Team (ISWT), a portion
of the total scientific observing time, the guaranteed time, will be
used for their Core Programme observations (Figure~\ref{cw:programme}).

\begin{table}[hp]
  \begin{center}
    \caption{Core Programme elements and allocated observing time during first
    year of operations.}\vspace{1em}
    \renewcommand{\arraystretch}{1.2}
    \begin{tabular}[h]{clc}
      \hline
      Priority & Core Programme &  Exposure\\   
      & Element & (10$^6$ s)\\
      \hline
      1& Deep exposure of &  \\
       & the Galactic central & 4.3 \\
       & radian (GCDE) & \\ \hline
      2  & Weekly scans of & \\
      & the Galactic plane (GPS) & 2.3 \\ \hline
      3  & Pointed observations: &  \\
         & $\bullet$ {\em transient events (TOO)} & {\em 1.7} \\
	 & $\bullet$ {\em Vela region}            & {\em 1.0} \\
	 & {\em total pointed observations} & 2.7 \\
      \hline
      Total & & 9.3 \\
      \hline \\
      \end{tabular}
    \label{cw:cp}
  \end{center}
\end{table}

The guaranteed time amounts
during the first year of nominal operations to 35\% (or $\sim$9.3$\times$10$^6$ s)
of the total annual observing time ($\sim$26.6$\times$10$^{6}$s)
\footnote{Note, that these numbers may decrease slightly as time for
instrument calibration observations during nominal operations needs to
be allocated}. This share
decreases to 30\% (year 2) and 25\% (to be confirmed, during extended mission).

The Core Programme \citep{gehrels97}, \citep{winkler99} will 
consist of three elements as shown in Table~\ref{cw:cp}.
The data from the Core Programme observations 
belong to the ISWT for the usual proprietary period of one 
year, after which the data will become public (see Section 4).

\section{Science Objectives of the Core Programme}

\subsection{Deep Exposure of the Galactic Central Radian (GCDE)}

The GCDE 
is driven by the following 
objectives: 
\begin{itemize}
\item mapping line emission from nucleosynthesis radioisotopes 
(e.g. $^{26}$Al, $^{44}$Ti, 511 keV), 
\item mapping continuum emission of the Galactic 
ridge, and 
\item performing deep imaging and spectroscopic studies of the 
central region of the Galaxy. 
\end{itemize}
Several interesting emission regions 
in or near the Galactic plane have been identified using CGRO OSSE and 
COMPTEL: these include the $^{26}$Al (7$\times$10$^5$ years half life) mapping the 
sites of nucleosynthesis in the past million years in the Galaxy 
\citep{diehl95}, and $^{44}$Ti (half life $\sim$60 years) 
which has been detected by COMPTEL from the Cas A SNR 
\citep{iyudin94} and more recently from a region coincident with
a new SNR in Vela, detected by ROSAT \citep{iyudin98}. 
OSSE mapping of the 
positron - electron annihilation radiation at 511 keV shows a 
central bulge, emission in the Galactic plane and an enhancement of 
extension of emission at positive latitudes above the Galactic 
centre \citep{purcell97}. Other isotopes such as $^{60}$Fe 
produce lines which could be detected by INTEGRAL. 
The origin of the clumpy structure of the COMPTEL observed 
$^{26}$Al maps and the $^{44}$Ti emission from hidden supernovae (''hot spots'')
are key targets of INTEGRAL research.

\begin{figure*}[htp]
\centering
\includegraphics[width=0.7\linewidth]{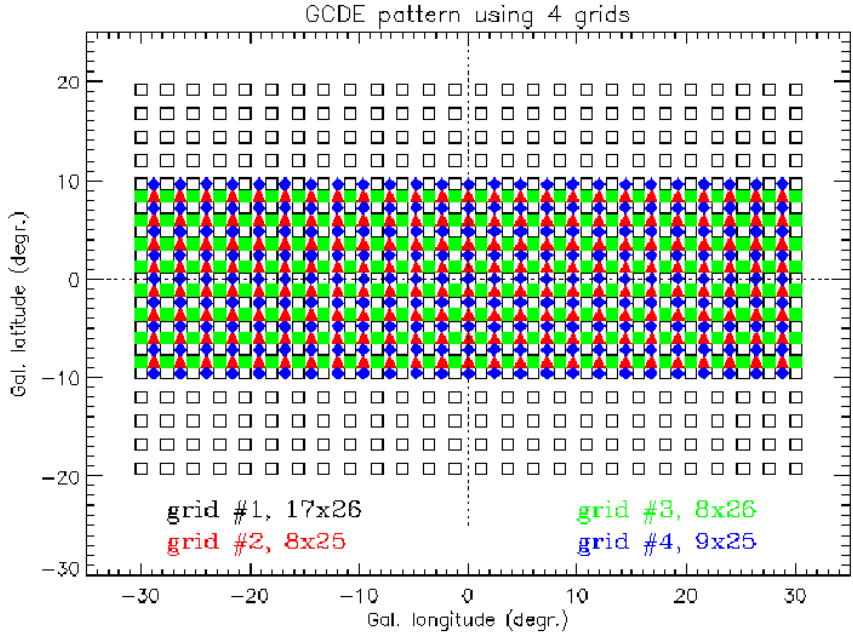}
\caption{Pointing grids for the GCDE exposure. The exposure per pointing
(separated by 2.4$^{\circ}$) is 30 minutes. The area (4 grids) will be
scanned twice per year. Grid 1: 442 pointings (open squares); Grid 2: 200 
pointings (triangles); Grid 3: 208 pointings (filled squares); 
Grid 4: 225 pointings (diamonds).}
\label{cw:gcde}
\end{figure*}

The INTEGRAL GCDE deep exposure will also study the continuum gamma-ray 
and hard X-ray emission from the Galactic plane. This ``galactic ridge'' 
is concentrated in a narrow band with a latitude extent of $\sim5^{\circ}$ 
and a longitude extent of $\pm40^{\circ}$ \citep{gehrels93}, 
\citep{valinia98}, \citep{strong99}.
The exact distribution and spectrum of the ridge emission is not 
well known. The origin is thought to be Bremsstrahlung from cosmic 
ray electrons, but this is also not fully established. INTEGRAL 
will be able to map the emission with high sensitivity and high 
angular resolution. This should allow the removal of the point-source 
origin of the emission so that the spectrum can be determined              
with high confidence. 
The GCDE will resolve isolated point sources with arcmin location and 
provide source spectra with high energy resolution. At least 90 
sources known as X- and gamma-ray emitters are contained in the 
region at $\sim10$ keV and GRANAT/SIGMA, BeppoSAX and other earlier missions 
have found this region to be filled with a number of highly 
variable and transient sources \citep{vargas97},
\citep{ubertini97}.
Many are thought to be compact objects in binary systems 
undergoing dynamic accretion. However, even for some of the 
brightest sources in the region (1E 1740.7$-$2942, GRS 1758$-$258) 
the detailed nature of the systems are not known. 
INTEGRAL will study the faintest sources with 
high angular resolution allowing multi-source monitoring 
within its wide field of view during single pointings. 
Also interesting will be searches for gamma-ray emission from 
SgrA$^*$, at the Centre of the Galaxy 
\citep{sunyaev93}.

The central radian of the Galaxy will be observed using a set of 
4 rectangular 
pointing grids each with a pitch of 2.4$^{\circ}$. The grids are shifted
with respect to each other by 1.2$^{\circ}$ (Figure~\ref{cw:gcde}).
The outer
grid covers the celestial region between $-$~30$^{\circ}$ $\le$ l $\le$ 
$+$~30$^{\circ}$ and $-$~20$^{\circ}$ $\le$ b $\le$ $+$~20$^{\circ}$, while the
three inner grids extend to galactic latitude $\pm$10$^{\circ}$ only.
This choice allows to obtain an exposure ratio of 4:1 of the inner 
($\le$ 10$^{\circ}$) part of the Galactic plane compared to the
outer ($\le$ 20$^{\circ}$) part. The entire
GCDE grid system is therefore optimized to minimise SPI imaging artefacts
and to optimize IBIS' sensitivity to point sources predominantly
located close to the Galactic plane. 
With an assumed 200 s slew duration for 2.4$^{\circ}$ and 1800 s 
exposure per point, about 2.15$\times$10$^6$ s will be required to 
scan all grids once. 
Within the total time allocation for the GCDE, 
all four grids will be scanned 2 times per year.

\subsection{Weekly scans of the Galactic plane (GPS)} 

The scanning of the Galactic plane (GPS) will 
be mainly done for two reasons: 
the most important one is to provide frequent monitoring of the plane in 
order to detect transient sources because the gamma-ray sky in the INTEGRAL 
energy range is dominated by the extreme variability of many sources. The 
scans would find sources in high state (outburst) which warrant possible 
scientifically important follow-up observations (Target of Opportunity [TOO] 
observation, see Section 2.3). The second reason is to build up 
time resolved maps of the Galactic plane in 
continuum and diffuse line emission such as $^{26}$Al, $^{44}$Ti and 511 keV with modest 
exposure. 
CGRO and SIGMA have been detecting Galactic compact sources of 
several different categories/groups which include 
X-ray binaries (e.g. X-ray novae, Be binary pulsars) and in particular
superluminal sources (GRS 1915$+$105 and GRO J1655$-$40). 
The occurrence rates for events that INTEGRAL can observe 
is about 2 events/year for 
each of these classes, where pointing constraints 
due to the fixed solar arrays have been taken into account. The important 
time scales for the transient outbursts vary significantly from class to 
class and from event to event, but a typical duration of an event is 1 - 2 
weeks and a typical variability time scale is of the order of 1 day. 

\begin{figure*}[htp]
\centering
\includegraphics[width=0.8\linewidth]{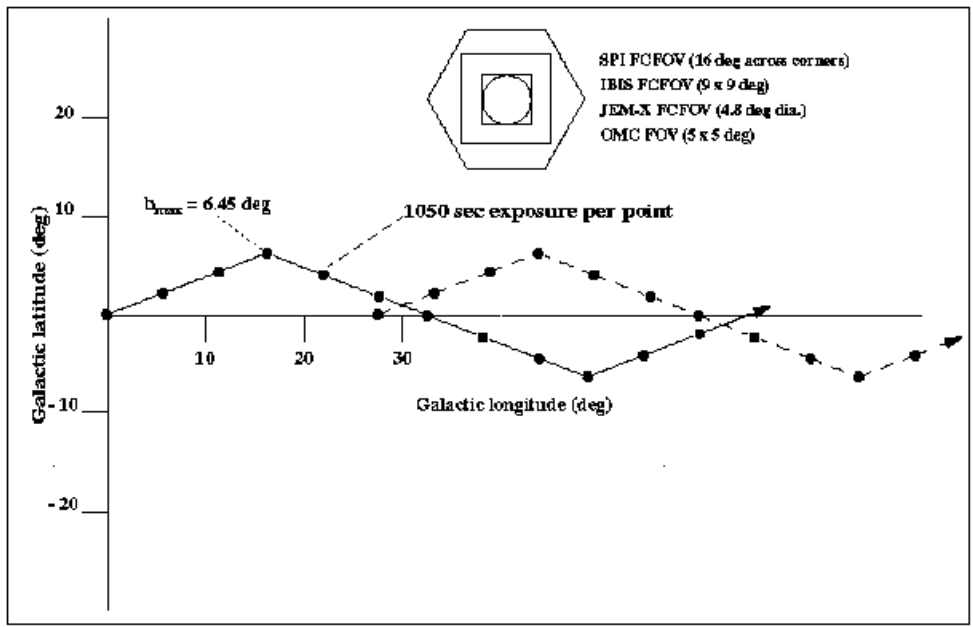}
\caption{Schematic view of two consecutive GPS scans. 
The angular distance 
between subsequent exposures (1050 s each) along the scan path is 6$^{\circ}$. 
The scan will be performed as a sawtooth 
with inclination of 21$^{\circ}$ with respect to the Galactic plane, 
each subsequent scan is shifted by 27.5$^{\circ}$ in galactic longitude.}
\label{cw:gps}
\end{figure*}

\begin{figure*}[htp]
\centering
\includegraphics[width=0.8\linewidth]{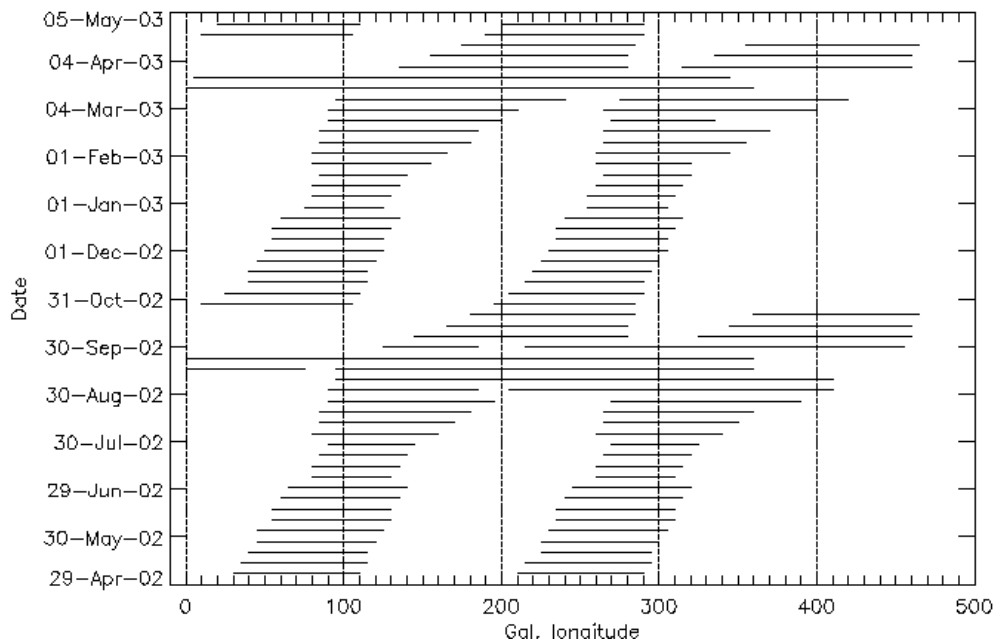}
\caption{Visibility of a b = $\pm$10$^{\circ}$ - wide strip used for
GPS scans centred on
the Galactic plane during the 
first year of operations assuming a 22 April 2002 launch date. The extent of
the visible part of the Galactic plane is shown by the solid lines.}
\label{cw:visib}
\end{figure*}

The scans will be 
performed once a week by performing a ``slew - and stare'' manoeuvre 
(Figure ~\ref{cw:gps}) 
of the spacecraft along the visible (accessible) part of the 
Galactic plane (Figure~\ref{cw:visib}) with latitude extent $\pm$10$^{\circ}$.
The accessible part of the Galactic plane depends on viewing constraints 
including the solar aspect angle (40$^{\circ}$ during first two years outside
eclipse seasons, 30$^{\circ}$ during first two years during
eclipse seasons, and
30$^{\circ}$ during extended mission phase) due to the fixed solar arrays, 
and on the season of the year.
 
The angular distance 
between subsequent exposures (1050 s each) along the scan path is 6$^{\circ}$. 
The scan will be performed as a sawtooth 
with inclination of 21$^{\circ}$ with respect to the Galactic plane, 
each subsequent scan is shifted by 27.5$^{\circ}$ in galactic longitude. 
The visibility (Figure~\ref{cw:visib}) of the Galactic plane depends on the annual season. On average
a strip of width b = $\pm$10$^{\circ}$ has a longitude 
extent of 178.4$^{\circ}$ (ranging from
100$^{\circ}$ to 350$^{\circ}$) and requires on average 12.3 hours of scanning
(ranging from 6.8 hours to 23.6 hours).

\subsection{Pointed observations}

Remaining Core Programme observing time (2.7$\times$10$^6$ s,
Table~\ref{cw:cp}) not spent on GPS and GCDE will 
be devoted to dedicated pointings on individual sources, which have been
defined as:
\begin{itemize}
\item transient events
\begin{itemize}
\item unknown transient events: local group SNII, SNIa,
classical novae and previously unknown X-/$\gamma$ transients, 
\item known transient events: GRS~1915$+$105, GRO~J1655, 1E~1740$-$2942, 
Cyg~X$-$1, Cyg~X$-$3, GX~339$-$4 and Mk~501, and,
\end{itemize}
\item Vela region, target pointing at (l,b) = (265$^{\circ}$,0$^{\circ}$). 
\end{itemize}

\subsubsection{Transient events}
Core Programme observing time of $\sim$1.7$\times$10$^6$ s will
be set aside to be able to perform scientifically important 
follow-up observations on those events ("TOOs") which have either been 
detected during INTEGRAL's
GPS and GCDE observations or where the trigger has been observed 
by optical telescopes 
(SN and Novae) or observations at other wavelengths.
Core Programme observation time will be used on 
any of the known and unknown sources listed above, 
whatever TOO event comes first.

\begin{figure*}[htp]
\centering
\includegraphics[width=0.8\linewidth]{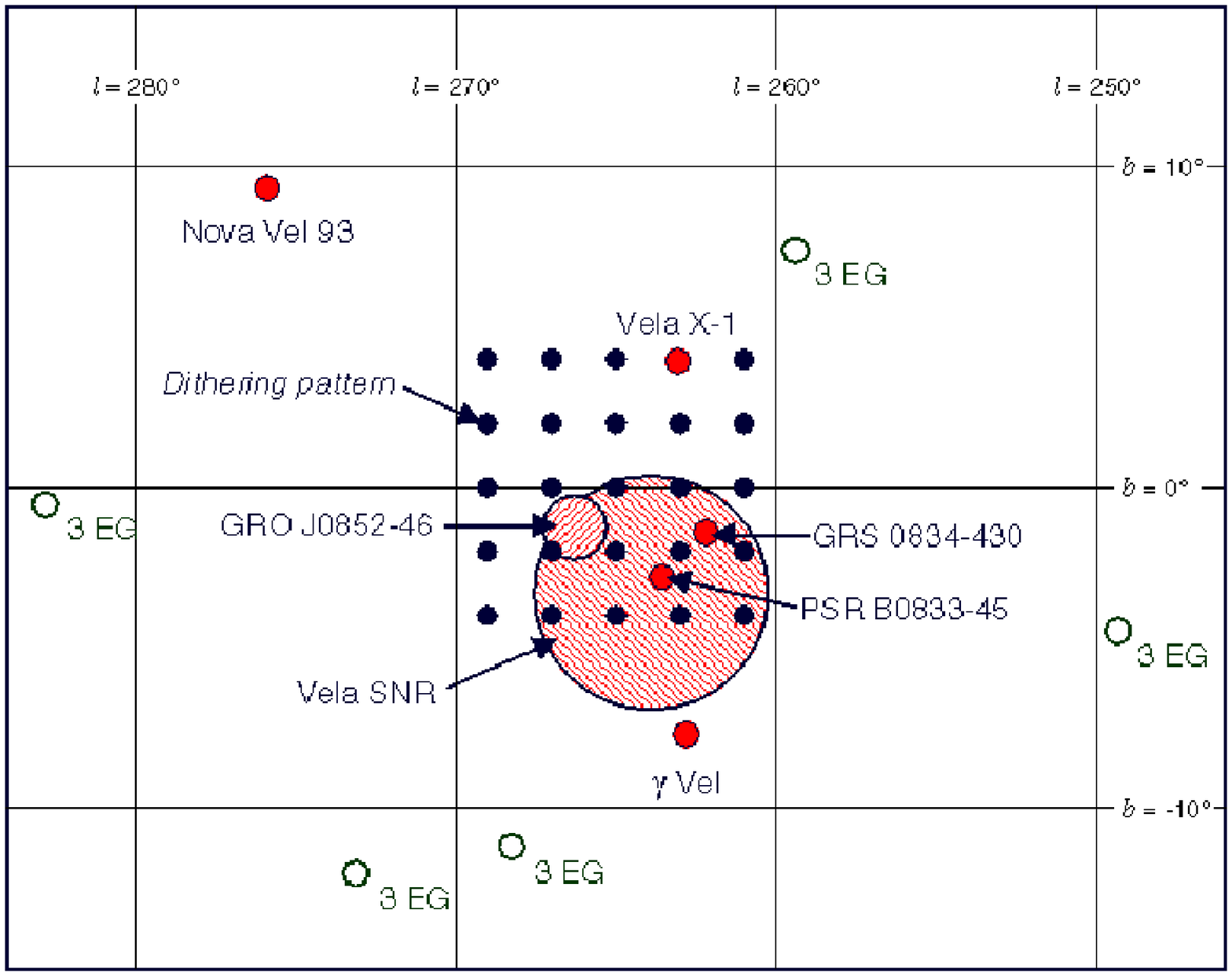}
\caption{Finding chart of the target sources to be scrutinized during the
proposed INTEGRAL survey of the Vela region.}
\label{cw:vela}
\end{figure*}

The greatest advantage of INTEGRAL, as compared to other high 
energy missions, is its high spectral resolution and 
high sensitivity at energies above 100 keV. 
The strongest point for 
INTEGRAL to response to a TOO call is the potential to detect 
transient high energy emission features (e.g. 511 keV, its scattering feature 
at 170 keV; 480 keV ($^7$Li); MeV bump; flares from Blazars).
The high energy cutoff or spectral break as a function of time may 
provide critical information to our understanding of the emission 
mechanism and system parameters. 
Concerning the high energy continuum, several pointings for each flare or 
outburst are 
required to achieve meaningful science return since the 
spectral evolution instead of simply the spectral shape is the 
main objective here. Separation between the exposures depends on 
the time scale of the event.

The chance that a supernova of type Ia occurs with detectable 
gamma-ray line emission from the 
$^{56}$Ni~$\rightarrow$~$^{56}$Co~$\rightarrow$~$^{56}$Fe 
decay chain in the INTEGRAL life-time, is promising. 

The maximum distance for a SNIa, to be detected by SPI, will depend
on the line broadening: 
If an expansion velocity of 10,000 km/s is assumed, the SPI 
sensitivity limit for detecting the 847 keV $^{56}$Co-line is 
about 3 $\times$ 10$^{-5}$ cm$^{-2}$ s$^{-1}$. 
For a $^{56}$Ni-yield of the supernova of 1 M$_o$, 
this results in a detectability up to a distance of 15.5 Mpc. 
Within this distance, one type Ia supernova can be 
expected every two or three years \citep{gehrels87}.
  
The chance to detect gamma-ray line emission from a nearby type 
II supernova is much smaller, because of the 
lower $^{56}$Ni-yield and the long obscuration 
time of type II supernovae to gamma-rays. 

Radioactive elements are being synthesized during a Nova explosion, 
yet to be detected through gamma-ray line emissions.
Short lived beta unstable nuclei may be detected via the 
511 keV annihilation emission. Today's most interesting 
gamma-ray lines are 1275 keV ($^{22}$Na, t = 3.8 y) and 478 keV 
($^7$Be $\rightarrow$ $^7$Li, t = 53 d).\\
{\em 1275 keV:} Recent results on two near novae 
(Nova Her 1991, Nova Cyg 1992) as observed by CGRO 
established 2$\sigma$ upper limits of $\sim$ 2 to 3 $\times$ 
10$^{-5}$ cm$^{-2}$ s$^{-1}$). 
SPI should be able to achieve a positive detection on similar events.\\
{\em 478 keV:} Assuming an ejected mass of 10$^{-5}$ M$_o$, 
SPI should detect 478 keV emission for close ($\sim$ 500 pc) 
novae just after outburst. 
INTEGRAL should provide evidence whether C-O novae do produce $^7$Li, 
which is under abundant in standard big bang 
nucleosynthesis as compared to solar 
abundance by more than an order of magnitude \citep{hernanz96}, 
\citep{hernanz99}.

\subsubsection{Vela region}

Finally, the remaining Core Programme observing time allocated to
pointed observations ($\sim$ 1.0$\times$10$^6$s, see Table~\ref{cw:cp}) will
be used for an observational survey of the Vela region, 
using the 5$\times$5 dither pattern centred on (l,b) = (265$^{\circ}$,0$^{\circ}$).
This region is a showcase for Galactic sources active in the INTEGRAL energy
domain, including massive stars, historical SNe, young SNR's, spin powered
pulsars. Furthermore, the survey intends to depict and study the diffuse
1.8 MeV interstellar source(s) as well as few unidentified EGRET (EG) sources
(Figure~\ref{cw:vela}).

\section{Exposure Times and Sensitivities}

\subsection{GCDE}

The area covered by the GCDE 
will receive a net exposure (i.e. excluding slews) 
of $\sim$ 4$\times$10$^6$ s per year 
which will give SPI a 3$\sigma$ sensitivity of 
$\sim$3.5$\times$10$^{-6}$ cm$^{-2}$s${^-1}$ for narrow 
lines in the 100 keV - 2 MeV region, sufficient for mapping and for 
detailed line shape studies at 1809 keV \citep{gehrels96}
of the bright $^{26}$Al ``hot spots'' ($\sim$3$\times$10$^{-5}$
cm$^{-2}$s$^{-1}$) 
as detected by COMPTEL.
The continuum sensitivities (3$\sigma$) in the 100 keV to 1 MeV range would 
be 
1$\times$10$^{-6}$ to 8$\times$10$^{-8}$ cm$^{-2}$ s$^{-1}$ keV$^{-1}$ 
(SPI) and 
(2.5 to 3)$\times$10$^{-7}$ cm$^{-2}$ s$^{-1}$ keV$^{-1}$
(IBIS) while JEM-X would achieve $\sim$3.5$\times$10$^{-6}$ 
cm$^{-2}$ s$^{-1}$ keV$^{-1}$
in the 3 to 30 keV range.

\begin{figure*}[htp]
\centering
\includegraphics[width=0.8\linewidth]{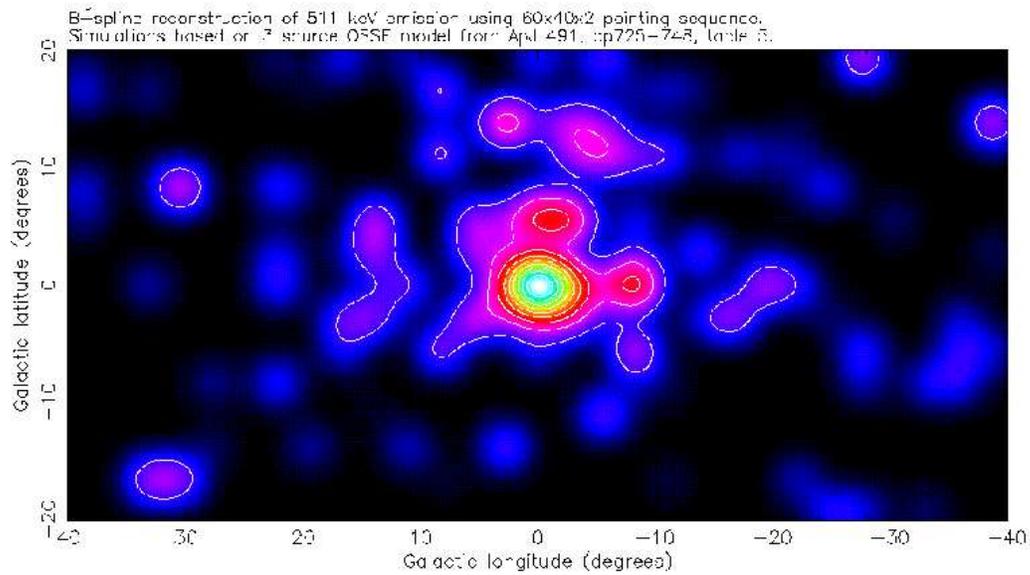}
\caption{Simulated image of a possible 511 keV emission model for GCDE (see
text), from \citet{jean}.}
\label{cw:sim2}
\end{figure*}

In Figure~\ref{cw:sim2} we show a SPI simulation \citep{jean} 
of the GCDE exposure at 511 keV taking the OSSE 3-source model 
(extended feature, point source at Galactic centre, 
emission at northern latitude) according to
\citet{purcell97} as input model into account.

\subsection{GPS}

\begin{figure}[htp]
\centering
\includegraphics[width=1.2\linewidth]{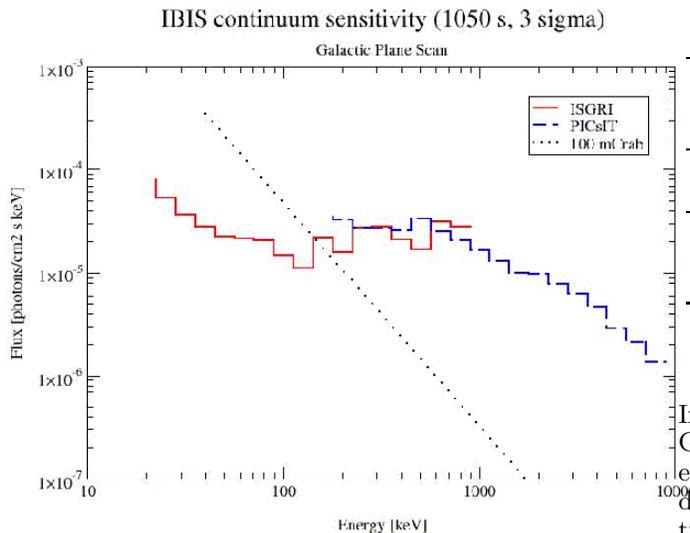}
\caption{IBIS continuum 3$\sigma$ sensitivity for a single point on a GPS sawtooth
scan (1050 s exposure).}
\label{cw:ibis}
\end{figure}

The IBIS continuum 3$\sigma$ sensitivity for a single point on a GPS sawtooth
scan (1050 s exposure, Figure~\ref{cw:gps}) is shown in Figure~\ref{cw:ibis}.
This will correspond to a 3$\sigma$ sensitivity of 
1.5$\times$10$^{-5}$ cm$^{-2}$ s$^{-1}$ keV$^{-1}$ @ 100 keV and 
$\sim$1.9$\times$10$^{-5}$ cm$^{-2}$ s$^{-1}$ keV$^{-1}$ @ 1 MeV.
For SPI a 1050 sec exposure would result in a 3$\sigma$ sensitivity
of 4.6$\times$10$^{-6}$ cm$^{-2}$ s$^{-1}$ keV$^{-1}$ @ 1 MeV (continuum)
and 1.6$\times$10$^{-4}$ cm$^{-2}$ s$^{-1}$ @ 1 MeV (narrow line).

Due to the large coded field-of-views for SPI (FCFOV 16$^{\circ}$ across)
and IBIS (FCFOV 9$^{\circ}$ $\times$ 9$^{\circ}$), a single ``staring'' 
point on a GPS scan will not only receive the nominal exposure per 
pointing (1050 s), but a larger amount given by the larger partially coded FOV:
While this exposure point is 
still located within the partially coded 
field-of-view (PCFOV, coverage to zero response), 
additional exposure will be collected as the spacecraft 
(i.e. the co-aligned instruments) is pointing at close neighbouring 
positions of that single exposure point (see Figure ~\ref{cw:gps}).

For SPI and IBIS, a single ``staring'' point will therefore 
receive a total exposure of 3465 s (SPI), and 3255 s (IBIS) on each scan, 
while JEM-X and OMC -- because
of the smaller FOV's -- will accumulate 1050 s per point/scan.

The 3$\sigma$ sensitivities for the high energy instruments 
which can be obtained for one ``staring'' point during a single scan, 
and for the same ``staring'' point for all scans during one year 
are listed in Table~\ref{cw:sens}. The exposure values for OMC are equivalent
to the JEM-X values. 
The expected sensitivities will allow detection of transient sources 
throughout the Galaxy.

\begin{table}[htp]
  \begin{center}
    \caption{GPS: sensitivities per exposure point along sawtooth.
    (1): Exposure within FCFOV per single pointing. (2): Exposure within
    FCFOV per pointing per year. (3): Units are cm$^{-2}$ s$^{-1}$ keV$^{-1}$.}
    \vspace{1em}
    \renewcommand{\arraystretch}{1.0}
    \begin{tabular}[h]{lll}
      \hline
      Instrument& Exposure$^{(1)}$
      & 3$\sigma$ sensitivity$^{(3)}$ \\
      &  (s) & @ energy (MeV) \\
      
      \hline
      JEM-X & 1050 & 4.0 $\times$10$^{-4}$ @ 0.006 \\
      IBIS  & 3255 & 8.8 $\times$10$^{-6}$ @ 0.1 \\
      SPI   & 3465 & 2.5 $\times$10$^{-6}$ @ 1 \\
      \hline
      & Exposure$^{(2)}$   &   \\
      & (s)  & \\
      \hline
      JEM-X & 26,900 & 7.9 $\times$10$^{-5}$ @ 0.006\\
      IBIS  & 83,500 & 1.7 $\times$10$^{-6}$ @ 0.1 \\
      SPI   & 88,900 & 4.9 $\times$10$^{-7}$ @ 1 \\
      \hline \\
      \end{tabular}
    \label{cw:sens}
  \end{center}
\end{table}

\begin{figure*}[htp]
\centering
\includegraphics[width=0.8\linewidth]{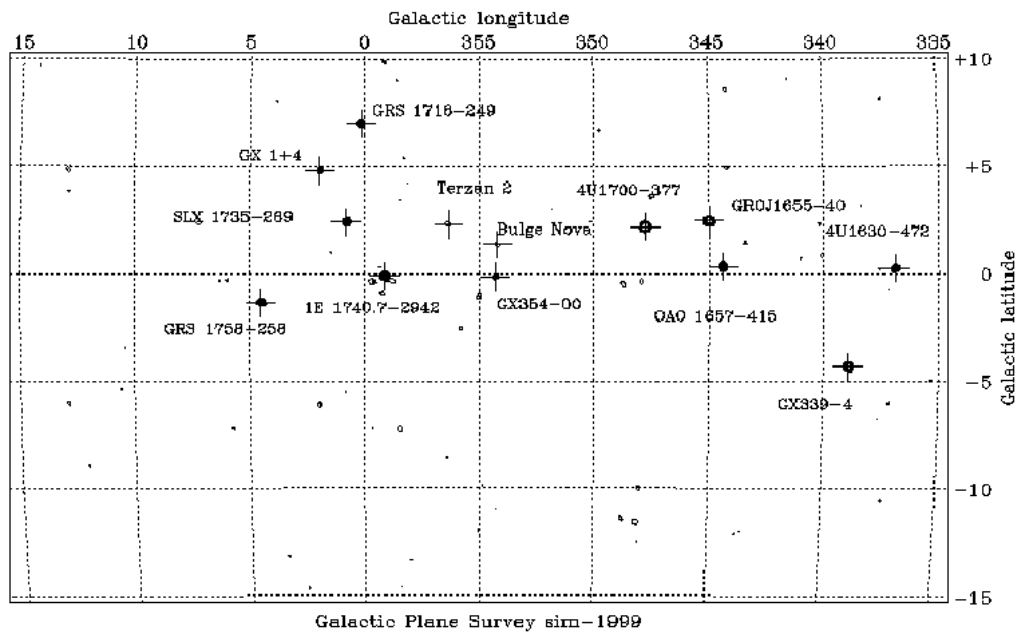}
\caption{Contour level image of the 50-150 keV reconstructed sky region
of a simulated GPS scan with IBIS. Levels start from 2.5$\sigma$, spaced by
1$\sigma$. All marked sources are detected at more than 4.5$\sigma$ 
\citep{goldwurm}.}
\label{cw:sim}
\end{figure*}

In Figure~\ref{cw:sim} we show IBIS simulation results of one GPS scan across
the Galactic Centre \citep{goldwurm}. Several known sources can be easily
detected at levels above 10$\sigma$. Furthermore, simulation of a 1 Crab
transient results in a 90\% error radius less than 10 arcseconds providing
accurate data for optical/radio follow-up observations. Finally, a faint source
(30 mCrab) located in a crowded region would be detected at the 6$\sigma$ level.

\subsection{Pointed observations}

\subsubsection{Follow-up observations of unknown TOO's}
Supernovae and Novae observations would be possibly done repeatedly
at various times after onset, to study, for example, early observations 
of radioactivity from Co breaking through the SN envelope and 
late observations of $^{44}$Ti. The typical total exposure would amount
to $\sim$ 1 - 2 $\times$10$^6$ s.

\subsubsection {Follow-up observations of known TOO's}
These sources also require a number of repeated observations after alert
on intervals ranging from days to weeks. Typical total exposure per source
would amount to $\sim$ 3$\times$10$^5$ s with an IBIS sensitivity
of $\sim$9$\times$10$^{-7}$ cm$^{-2}$ s$^{-1}$ keV$^{-1}$ (0.07 mCrab) 
at 100 keV. 
These observations will be
triggered if the source deviates significantly from a ''normal'' intensity
level. These would be, for example for GRS1915$+$105, a flux exceeding
500 mCrab @ 10 keV, or 300 mCrab @ 100 keV for GRO~J1655, easily detectable
by a single GPS exposure.

\subsubsection{Survey of the Vela region}
Studies of points sources and 
nucleosynthesis studies of the Vela regions require 
typically $\sim$10$^6$ s.  
The continuum sensitivities (3$\sigma$) obtainable
during that observation in the 100 keV to 1 MeV range would 
be 
2$\times$10$^{-6}$ to 1.5$\times$10$^{-7}$ cm$^{-2}$ s$^{-1}$ keV$^{-1}$ 
(SPI) and 
(5 to 6)$\times$10$^{-7}$ cm$^{-2}$ s$^{-1}$ keV$^{-1}$
(IBIS) while JEM-X would achieve $\sim$7$\times$10$^{-6}$ 
cm$^{-2}$ s$^{-1}$ keV$^{-1}$
in the 3 to 30 keV range. For line studies SPI will achieve a 3$\sigma$
sensitivity of $\sim$7$\times$10$^{-6}$ 
cm$^{-2}$ s$^{-1}$ in the 100 keV to 2 MeV region.

\section{Data analysis and data rights}

Data from the Core Programme will be used by the ISWT as a common
scientific database. This database will be analysed according to research
categories (e.g. diffuse galactic emission) and topics (e.g. 511 keV line
profiles). The topics/scientists involved in that research do match the
agreed share of Core Programme data within the ISWT.
The data are property of the ISWT for the duration of one year. Then
they will become publicly available through the archive. The same rules
do apply for data obtained during the open time (General Programme).

The specification\footnote{The final list of targets will be
made available in the INTEGRAL AO-1 documentation.} 
of the known and unknown transients (see above) serves
the purpose to clearly notify the scientific community at large on 
the CP guaranteed targets. Note, however,
that due to the random nature of transient events and the limited amount
of CP allocated time, it is possible that not all pointed TOO follow-up
observations will be performed within the CP allocated time. 
Therefore open time proposals for any of those TOO sources as listed above 
are not a priori prohibited and should differ in scientific objectives
and instrument modes but run
a smaller chance of being executed than any open time TOO 
proposals made for transient events
not listed in the CP. The Vela region is reserved as a guarenteed
time target.

In case TOO follow-up observations will be performed for (exciting)
scientific reasons,
and the required time would exceed the total allocated ceiling, and no
open time proposals would exist, then these observations would be performed,
but the data would be made publicly available immediately. However, the ISWT
will keep some flexibility in the \underline{overall} allocation
of its CP time in case a major TOO event (e.g. close-by SN) would occur and
not sufficient time would be available anymore. In this case the ISWT
will review the CP priorities and, if available, re-assign GCDE and GPS
time for this follow-up observation. In case of low TOO activity, more time
could be spend on the Vela region observation.

\section*{Acknowledgments}

The INTEGRAL Core Programme has been defined by the Core Programme Working Group
(N. Gehrels, N. Lund, P. Ubertini, V. Sch\"onfelder, C. Winkler) who has been
charged by the ISWT with this task. The fundamental support of the
instrument teams, in particular through extensive simulation studies,
and of the INTEGRAL Mission Scientists is gratefully acknowledged.

\end{document}